\documentclass[twocolumn,prb,amsmath,amssymb,aps]{revtex4-2}
\usepackage{dcolumn}
\usepackage{autobreak}
\usepackage{braket}
\usepackage{graphicx}
\usepackage{booktabs}
\usepackage{bm}
\usepackage[colorlinks, linkcolor=blue, anchorcolor=blue, citecolor=blue]{hyperref}

\makeatletter

\newcommand{\Rmnum}[1]{\expandafter\@slowromancap\romannumeral #1@}
\newcommand{\SRO}{Sr$_2$RuO$_4$}

\newcommand{\bsigma}{\bar{\sigma}}

\makeatother

\usepackage{mynote}

\begin{document}
\title{Impact of random impurities on the anomalous Hall effect in chiral superconductors}
	\author{Hao-Tian Liu}
	\address{Shenzhen Institute for Quantum Science and Engineering, Southern University of Science and Technology, Shenzhen 518055, Guangdong, China}
	\address{International Quantum Academy, Shenzhen 518048, China}
	\address{Guangdong Provincial Key Laboratory of Quantum Science and Engineering, Southern University of Science and Technology, Shenzhen 518055, China}
		\author{Weipeng Chen}
	\address{Shenzhen Institute for Quantum Science and Engineering, Southern University of Science and Technology, Shenzhen 518055, Guangdong, China}
	\address{International Quantum Academy, Shenzhen 518048, China}
	\address{Guangdong Provincial Key Laboratory of Quantum Science and Engineering, Southern University of Science and Technology, Shenzhen 518055, China}
	\author{Wen Huang}
	\email{huangw3@sustech.edu.cn}
	\address{Shenzhen Institute for Quantum Science and Engineering, Southern University of Science and Technology, Shenzhen 518055, Guangdong, China}
	\address{International Quantum Academy, Shenzhen 518048, China}
	\address{Guangdong Provincial Key Laboratory of Quantum Science and Engineering, Southern University of Science and Technology, Shenzhen 518055, China}
	
\date{\today}

\begin{abstract}
The anomalous Hall effect and the closely related polar Kerr effect are among the most direct evidence of chiral Cooper pairing in some superconductors. While it has been known that disorder or multiband pairing is typically needed for these effects to manifest, there is a lack of direct real-space investigation with regard to how disorder impacts the Hall response in both single-band and multiband chiral superconductors. On the basis of chiral superconducting models often adopted for \SRO, we study in this work the anomalous Hall effect in the presence of random non-magnetic impurities on real-space lattices. The single-band chiral p-wave ($p_x+ip_y$) calculation qualitatively reproduces the Hall conductivity obtained in previous skew-scattering-type diagrammatic analyses, along with some quantitative difference originating primarily from contributions involving impurity-induced in-gap states. The non-p-wave chiral states, such as $d_{x^2-y^2}+id_{xy}$, generically exhibit finite Hall response in the presence of random impurities, in contrast to a conclusion drawn from the aforementioned diagrammatic study. In particular, while pointlike impurities appears to induce minuscule Hall conductivity in non-self-consistent calculations, self-consistency and finite-range impurity potentials can both lead to substantial Hall conductivity. On the other hand, the intrinsic Hall conductivity in multiband chiral superconductors, which is related to interband transitions, decreases parametrically as disorder suppresses the superconducting order parameter. In addition, we check that random impurities do not induce anomalous Hall effect in non-chiral but time-reversal symmetry breaking superconducting states the likes of $s+i d_{x^2-y^2}$ and $d_{x^2-y^2}+i g_{xy(x^2-y^2)}$. We briefly remark on the implications of our results for Kerr effect measurements. 
\end{abstract}

\maketitle
\section{Introduction}
Chiral superconductivity is characterized by a time-reversal-symmetry-breaking (TRSB) Cooper pairing that develops a spontaneous orbital angular momentum, such as $p+i p$ and $d+i d$ pairing~\cite{Kallin2016}. In analogy to what Landau levels do in quantum Hall insulators, such Cooper pair condensate may support Hall-like transport in the absence of external magnetic field~\cite{Volovik1988}, i.e.~an anomalous Hall effect. The corresponding Hall conductivity~\cite{footnote}, however, is absent in clean single-band chiral superconductors due to the Galilean invariance principle, which states that the center-of-mass motion of a Cooper pair under an external electric field is oblivious to the relative motion between the two paired electrons~\cite{Read2000}. From a semiclassical standpoint~\cite{Luttinger1954, Niu1999}, the absence of Hall effect may be related to the vanishing of the anomalous velocity the Bogoliubov quasiparticles acquire when subject to an electric field.

The Galilean invariance is no longer preserved if the underlying translational symmetry is broken~\cite{Goryo2008, Lutchyn2009, Sigrist2001}, or, if the Cooper pairing takes places in a system with multiple Bloch bands~\cite{Kallin2012, Gyorffy2012}. These two scenarios provide, respectively, extrinsic and intrinsic mechanisms to entangle the relative and center-of-mass motion of a Cooper pair. Both mechanisms may generate finite ac anomalous Hall conductivity in chiral superconductors, which can then be probed in optical polar Kerr measurements~\cite{Kapitulnik2009}. Finite Kerr rotation below the superconducting transition have been reported in a number of putative chiral superconductors, including \SRO~\cite{Xia2006}, UPt$_3$~\cite{Schemm2014}, URu$_2$Si$_2$~\cite{Schemm2015} and UTe$_2$~\cite{Hayes2021}.

\SRO~in particular has been in a state of much controversy. Despite multiple early observations pointing to chiral $p_x+i p_y$ pairing~\cite{Sigrist1998, Maeno1998, Duffy2000, Nelson2004}, this superconducting order has faced increasing scrutiny~\cite{Kirtley2007,Hicks2010,Brown2019,Chronister2021}. The polar Kerr effect, besides indicating TRSB, also places other stringent constraints on the nature of the superconducting state. Specifically, it was argued that, for \SRO, only chiral Cooper pairings and two non-unitary mixed-helical-p-wave pairings may exhibit anomalous Hall effect and hence the Kerr effect~\cite{Zhang2020,Huang2021,Huang2021CPB}. Other TRSB states, such as $s+i d_{x^2-y^2}$~\cite{Anderson2019} and $d_{x^2-y^2}+i g_{xy(x^2-y^2)}$~\cite{Kivelson2020, Ghosh2021}, do not support anomalous Hall response in the clean limit. Nonetheless, impurities, dislocations or other forms of disorder may incur local symmetry breakings favorable for the Hall effect to arise. In this work, we study the impact of random non-magnetic impurities on the Hall response in various TRSB superconductors, with a focus on the chiral superconducting states. 

\begin{table*}
    \centering
    \caption{Superconducting gap functions considered in our calculations for single-band models of \SRO.}.
    \begin{tabular}{c|c|c}
    \hline
Superconductivity & Representation & Basis function\\
\hline
$s$-wave	& $A_{1g}$ & 1 \\
$g_{xy(x^2-y^2)}$ & $A_{2g}$ & $\sin k_x\sin k_y(\cos k_x-\cos k_y)$ \\
$d_{x^2-y^2}$ & $B_{1g}$ &$\cos k_x-\cos k_y$ \\
$d_{xy}$ &$B_{2g}$ & $\sin k_x\sin k_y$ \\ 
$p_x+ip_y$  & $E_u$ & $(\sin k_x,\sin k_y)$ \\
\hline
\end{tabular}
\label{pairing}
\end{table*}

The effect of non-magnetic impurity scatterings in single-band chiral superconductors has in fact been studied by means of diagramatic analyses~\cite{Goryo2008, Lutchyn2009,Konig:17}. An interesting observation following from Goryo's analysis on the skew-scattering-type processes, is the vanishing of Hall conductivity in all higher angular momentum chiral states (non-p-wave, such as chiral $d_{x^2-y^2}+id_{xy}$) in continuum models~\cite{Goryo2008}. These analyses, however, have only considered $s$-wave scattering off pointlike impurities, and they did not include the contribution
from the possible subgap quasiparticle states formed around impurities~\cite{Okuno1999,Balatsky2006}. In a separate study of impurity-induced thermal Hall effect, Ngampruetikorn and Sauls~\cite{Ngampruetikorn:20} found that non-pointlike impurities with finite-range impurity potential can induce finite thermal Hall response in non-p-wave states. On the other hand, Li et.~al.~\cite{Li2020} studied the low-energy theories emerging from superlattices of pointlike impurities embedded in both p-wave and non-p-wave chiral states, and found that the resultant impurity bands are also able to support anomalous Hall effect. Furthermore, none of the aforementioned studies has taken into account the superconducting order parameter inhomogeneity in the presence of impurities. A goal of the present study is to examine, through real-space simulations of two-dimensional (2D) lattice models, how random impurities, both pointlike and non-pointlike, impact the (electric) Hall response. It is also worth noting that such real-space calculations by default account for higher order scattering diagrams not included in previous studies in Refs.~\onlinecite{Goryo2008, Lutchyn2009,Konig:17}.

Our calculations of single-band models on a square lattice will corroborate the previous diagramatic analyses, while also offer the following new results. 1), due to the influence of impurity-induced subgap states, the p-wave Hall conductivity around $\omega=2\Delta$ exhibits quantitative difference from previous results~\cite{Lutchyn2009}. 2), in contrast with Goryo's conclusion~\cite{Goryo2008} but consistent with Li et.~al.~\cite{Li2020}, pointlike impurities does induce finite Hall conductivity in the $d_{x^2-y^2}+i d_{xy}$ state. The conductivity is minuscule if the spatial variations of the superconducting order parameters around impurities are neglected (i.e. non-self-consistent calculations), but it becomes substantial with self-consistency. Therefore, to correctly understand the Hall response in chiral d-wave, it is important to account for the real-space superconducting inhomogeneity. 3), in agreement with Ngampruetikorn and Sauls~\cite{Ngampruetikorn:20}, non-pointlike finite-range impurities readily generate substantial Hall conductivity in the $d_{x^2-y^2}+i d_{xy}$ state. The latter two can be generalized to higher angular momentum chiral states where they apply. In addition, we also check that Hall conductivity vanishes in disordered non-chiral but TRSB states such as $s+i d_{x^2-y^2}$ and $d_{x^2-y^2}+i g_{xy(x^2-y^2)}$. 

The intrinsic anomalous Hall effect in multiband (multi-orbital) chiral superconductors is generated by virtual interband optical transitions~\cite{Kallin2012, Gyorffy2013}. Hence the corresponding Hall conductivity emerges at frequency windows matching certain band separation energies -- scales that are typically much larger than the superconducting gap. This intrinsic mechanism was proposed~\cite{Kallin2012,Gyorffy2012} to explained the Kerr rotation in very clean samples of \SRO~\cite{Xia2006} and UPt$_3$~\cite{Schemm2014}. However, the fate of this intrinsic Hall response against disorder is an interesting question not yet explored. This will constitute another theme of our study. Through self-consistent real-space modeling of a two-band chiral p-wave model with random impurities, we shall find that the intrinsic Hall conductivity follows the parametric disorder-suppression of the superconducting order parameters. This contrasts with the behavior of the intrinsic longitudinal conductivity, which originates from similar interband optical transitions but is independent of the superconducting pairing. 

The remaining of the paper is organized as follows. Section \ref{sec:formalism} introduces the formalism for evaluating Hall conductivity in real-space lattice models. Secs.~\ref{sec:singleband} and \ref{sec:twoband} present our numerical results and analyses for single-band and two-band models, respectively. For the single-band model, we show results of models separately with pointlike and non-pointlike impurities, and without and with order parameter self-consistency. The two-band results are obtained from self-consistent calculations, unless otherwise specified. Section \ref{sec:remarks} summarizes our main results and remarks on the implication for Kerr effect measurements.

\section{Hall conductivity in real-space calculations}
\label{sec:formalism}
For notational simplicity, we set $\hbar=c=e=k_B=1$ throughout the study. According to standard linear response theory, the ac conductivity in the clean limit and at zero-external wavevector is given by the Kubo formula and is related to the current-current correlation function~\cite{Kallin2012}: 
$\pi_{\mu\nu}(i\nu_m) = T/\sum_{\bk,\omega_n}\tr\big[\hV_{\mu,\bk}\mathcal{G}(\bk,\omega_n)\hV_{\nu,\bk}\mathcal{G}(\bk,\omega_n+\nu_m)\big]$. In this expression, $T$ stands for the temperature, $\bk$ labels the crystal momentum, $\nu_m$ is the bosonic Matsubara frequency $\nu_m=2m\pi T$ with integer $m$, $\omega_n$ is the fermionic Matsubara frequency $\omega_n=(2n+1)\pi T$ with integer $n$, $\hV_{\mu,\bk}$ is the velocity (current) operator, and $\mathcal{G}$ is the Matsubara Green's function associated with the (Bogoliubov-de Gennes) Hamiltonian $\hH$, expressed by a resolvent $\mathcal{G}(\bk,\omega_n)=[i\omega_n-\hH(\bk)]^{-1}$. The Hall conductivity is defined by the anti-symmetric part of the tensor: $\sigma_H(\omega)=\frac{i}{2\omega}[\pi_{xy}(\omega)-\pi_{yx}(\omega)]$. For computational convenience, one may express the Green's function in the spectral representation:
\begin{align}
    \mathcal{G}(\bk,\omega_n)=\sum_{m}\frac{\ket{m,\bk}\bra{m,\bk}}{i\omega_n-\varepsilon_{m,\bk}}
\end{align}
where $\ket{m,\bk}$ is the $m$-th eigenstate of $\hH(\bk)$ with energy $\varepsilon_{m,\bk}$. Sum over the Matsubara frequency and apply analytic continuation $i\omega_n \ra \omega+i\eta$, we arrive at the ac Hall conductivity:
%\begin{widetext}
\begin{align}\label{hallcond0}
    \sigma_H(\omega)&=\frac{i}{4N\omega}\sum_{\bk,m,n}\frac{f(\varepsilon_{n,\bk})-f(\varepsilon_{m,\bk})}{\omega+i\eta-\varepsilon_{m,\bk}+\varepsilon_{n,\bk}}\non\\
    &\times \big[V^{mn}_{x,\bk}V_{y,\bk}^{nm}-(x\leftrightarrow y)\big]
\end{align}
%\end{widetext}
where $V_{\mu,\bk}^{mn}=\bra{m,\bk}\hV_{\mu,\bk}\ket{n,\bk}$ is the matrix element of the velocity operator $\hV_{\mu,\bk}$, $f(\varepsilon)$ denotes the Fermi-Dirac function, and $N$ is the number of sites in the system. 

In a real-space formulation, the above momentum space construction is no longer applicable. The conductivity is now derived from the correlations of the velocity operators associated with the real-space Hamiltonian $\hV_{\mu}$. It can be shown to acquire the following form,
\begin{align}\label{hallcond}
    \sigma_H(\omega)&=\frac{i}{4N\omega}\sum_{m,n}\frac{f(\varepsilon_{n})-f(\varepsilon_{m})}{\omega+i\eta-\varepsilon_{m}+\varepsilon_{n}}\non\\
    &\times \big[V^{mn}_{x}V_{y}^{nm}-(x\leftrightarrow y)\big]
\end{align}
where $V_{\mu}^{mn}=\bra{m}\hV_{\mu}\ket{n}$ with $|m\rangle$  denoting the eigenstate wavefunction, and $\varepsilon_{m}$ is the eigen-energy of the corresponding Hamiltonian.

For illustration, we hereby derive the velocity operator for a single-orbital tight-binding model on a square lattice with only nearest neighbor hoppings. The corresponding Hamiltonian is giving by $\hH_{\textrm{TB}}=-\mu\sum_{\bi}\hcd_{\bi}\hc_{\bi}+t\sum_{\braket{\bi,\bj}}(\hcd_{\bi}\hc_{\bj}+\HC)$, where $\mu$ is the chemical potential, $\bi=(i_x,i_y)$ is the two-dimensional position index. The spin indices are suppressed here as we do not consider spin-orbit coupling. Following the standard Peierls substitution, in the presence of a vector potential $\bA(\br)$, the hopping integral $t_{\bi\bj}$ between any pairs of sites $\bi$ and $\bj$ is replaced by $t_{\bi\bj}e^{i \int_{\br_{\bi}}^{\br_{\bj}}\bA(\br')\ud \br'}$~\cite{Hofstadter1976}. The current operator $\hat{\bj_0}$ for the normal state is then obtained by taking a partial derivative with respect to $\bA(\br)$ and then set $\bA(\br)\equiv 0$, 
\begin{align}
    \hat{\bj}_0=\sum_{\braket{\bi,\bj}}\bR_{\bi\bj}(i t\hcd_{\bi}\hc_{\bj}+\HC),\quad  \bR_{\bi\bj}\equiv \br_{\bi}-\br_{\bj}
\end{align}
The velocity operator is thus $\hat{\boldsymbol{V}}_0=\hat{\bj}_0/e=\hat{\bj}_0$, with its component along the $\mu$-direction denoted by $\hV_{0,\mu}$. Specific to superconducting models, the velocity operator in the Nambu spinor basis $\Psi=\begin{pmatrix}\hc_{(1,1)},\cdots,\hc_{(N_x,N_y)},\hcd_{(1,1)},\cdots,\hcd_{(N_x,N_y)}\end{pmatrix}^t$ reads as follows,
\begin{align}
    V_{\mu}=\begin{pmatrix}
    V_{0,\mu} & 0\\
    0 & V_{0,\mu}
    \end{pmatrix} \,.
\end{align}
In our study, we model disorder on the square lattice by assigning a pointlike potential $V_{\imp}=u$ much larger than the bandwidth to certain fraction of randomly chosen sites, $\hH_{\imp}=\sum_{\bi}u\hcd_{\bi}\hc_{\bi}$. At times, we add additional potentials to the neighboring sites of each impurity, in order to simulate the effect of finite-range (non-pointlike) impurities. For any given impurity concentration $n_\text{imp}$, the Hall conductivity is given by the average over an ensemble of impurity configurations $\sigma_H=\braket{\sigma}_{\imp}$. The above procedures can be generalized to multi-orbital models with ease. 

Below, we present the results and analyses for both single-band and two-band models of \SRO, although the conclusions shall be applicable to other TRSB superconductors. The superconductivity in \SRO~is typically described on the basis of a square lattice model. For TRSB pairings we consider the following four that frequently appear in literature: $p_x+i p_y$, $d_{x^2-y^2}+i d_{xy}$, $s+i d_{x^2-y^2}$, $d_{x^2-y^2}+i g_{xy(x^2-y^2)}$. In conjunction with the this compound's $D_{4h}$ point group symmetry, the chiral p-wave belongs to the $E_u$ irreducible representation, while the latter three are referred to as $B_{1g}+i B_{2g}$, $A_{1g}+i B_{1g}$ and $B_{1g}+i A_{2g}$ states. 

%\end{widetext}
\section{Single-band models}
\label{sec:singleband}

\subsection{Non-self-consistent calculations}
\label{subsec:nSC}
As a starting point, let's consider non-self-consistent calculations in which we neglect the spatial variation of the superconducting order parameters around local impurities. For illustrative purpose, we adopt simple gap functions whose momentum-space forms are given in TABLE \ref{pairing}. In real-space lattice realization, we have
$\hat{\Delta}=\sum_{\bi,\bj}(\Delta_ {\bi \bj, \sigma\bsigma}\hcd _{\bi \sigma}\hcd _{\bj , \bsigma}+\HC)$, where $\sigma$ denotes the spin and $\bsigma=-\sigma$, and for the aforementioned various irreducible representations,
\begin{align}
	A_{1g}:\, &\Delta _{\bi \bi, \sigma\bsigma}=\Delta \,,\\
	A_{2g}:\, &\Delta_ {\bi,\bi+2\hat{x}\pm \hat{y},\sigma\bsigma}=\mp\frac{\Delta}{2},\quad \Delta_ {\bi ,\bi+\hat{x}\pm 2\hat{y},\sigma\bsigma}=\pm \frac{\Delta}{2} \,, \non\\
	&\Delta_{\bi,\bi-2\hat{x}\pm\hat{y},\sigma\bar{\sigma}}=\pm \frac{\Delta}{2},\quad \Delta_{\bi,\bi-\hat{x}\pm 2\hat{y},\sigma\bar{\sigma}}=\mp\frac{\Delta}{2}\,,\\
	B_{1g}:\, &\Delta_ {\bi,\bi\pm\hat{x},\sigma\bsigma}=-\Delta _{\bi,\bi\pm\hat{y},\sigma\bsigma}=\frac{\Delta}{2} \,,\\
	B_{2g}:\, &\Delta _{\bi , \bi +\hat{x}\pm\hat{y},\sigma\bsigma}=\mp \frac{\Delta}{4},\quad \Delta_{\bi,\bi-\hat{x}\pm\hat{y}}=\pm\frac{\Delta}{4} \,,\\
	E_u:\, &\Delta _{\bi , \bi \pm \hat{x},\sigma\bsigma}=\pm\frac{\Delta}{2},\quad \Delta_{\bi ,\bi \pm \hat{y},\sigma\bsigma}=\mp i\frac{\Delta}{2} \,.
\end{align}

Calculations were performed on lattices of size $80\times 80$ with periodic boundary condition in both directions, and at least $30$ samples were used for impurity ensemble average. The pairing amplitude is typically chosen to be of order $0.2t$ in order to avoid finite-size effects. We set $\mu = t$, which gives a Fermi level density of states $N_0 \sim 0.3/t$ per spin species. The electron scattering rate, i.e.~the inverse electron lifetime, is given by $\Gamma = \tau^{-1} \sim n_\text{imp}/\pi N_0$. For the level of impurity concentration employed  in most of our calculations, $n_\text{imp} \leq 4\%$, the normal state remains a good metal with $E_F \tau \sim t\tau \gg 1$. The choice of parameters also ensures that the system lies near the limit $\Delta \gg \Gamma$, where superconductivity is expected to remain robust. This will also be confirmed in our self-consistent calculations, in the latter part of this section.

\begin{figure}[htbp]
    \centering
    \includegraphics[width=8.5cm]{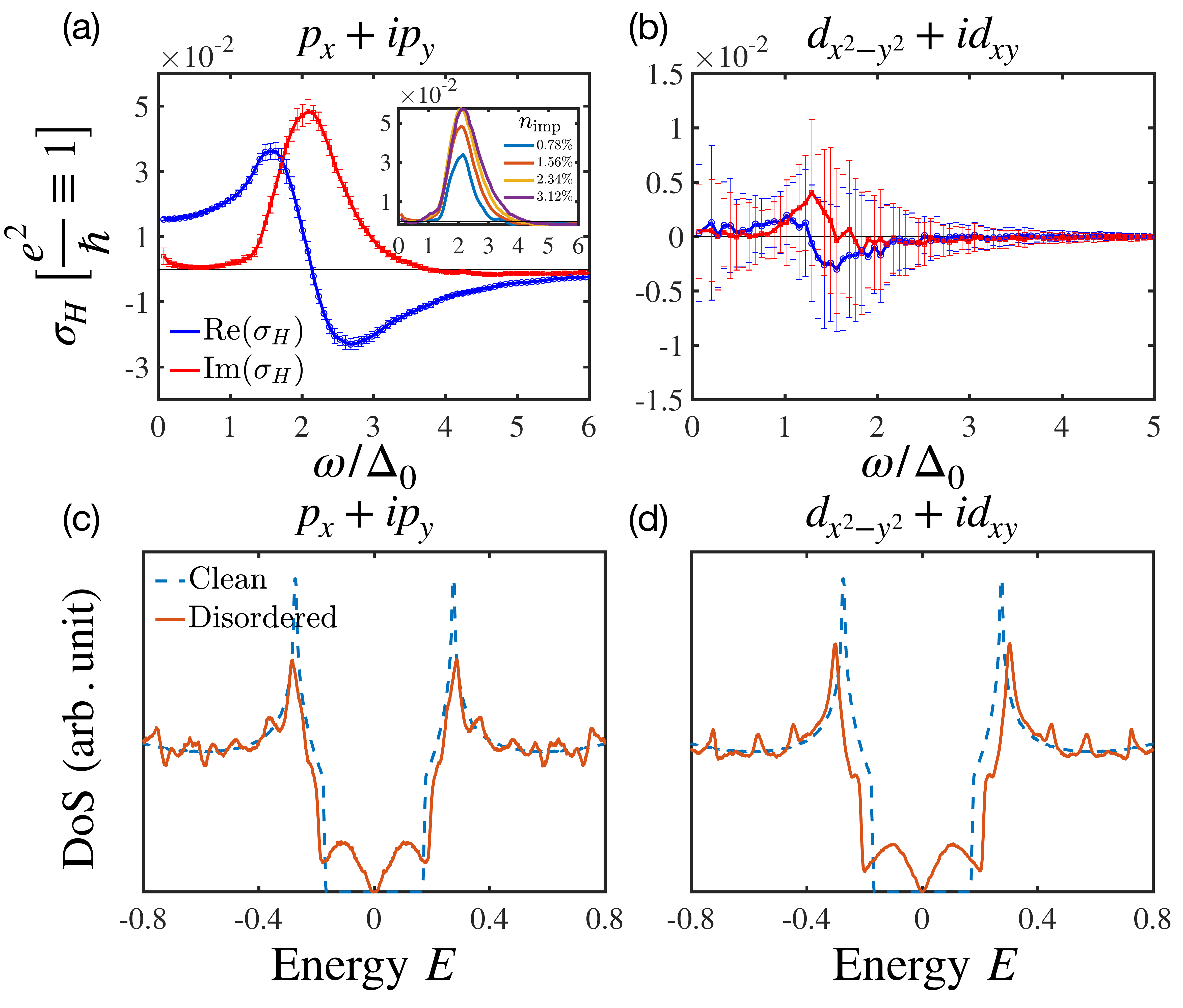}
    \caption{The zero-temperature Hall conductivity and the respective density of states for various single-band superconductors in 2D square lattices with randomly distributed pointlike impurities with concentration $n_{\imp}=1.56\%$. Through our calculations, $t=\mu=1$, $u=1000$ and $\eta=10^{-3}$. In (a) and (b), $\Delta_0$ is the quasiparticle excitation gap extracted respectively from the clean-limit DoS data in (c) and (d), with $\Delta_0=0.17t$ and $\Delta_0=0.18t$ respectively. The small upturn of $\text{Im}(\sigma_H)$ near zero frequency in (a) is a numerical artifact that can be eliminated by choosing sufficiently small $\eta$. The inset of (a) displays the imaginary part of the conductivity of the $p_x+ip_y$ state with different impurity concentrations. The set of data in (b) was obtained from an average over 90 random samples. The error bars reflect the scatter of data of different impurity configurations. For clarity, the error bars will be dropped in other figures.}
    \label{fig1}
\end{figure}

Representative numerical results for the zero-temperature ac Hall conductivity in the presence of pointlike impurities are shown in Fig.~\ref{fig1}. It can be seen that the Hall response varies drastically among the four superconducting states. Overall, the Hall conductivity is finite in the chiral states and vanishes in nonchiral states. At finite temperatures, the conductivity tracks the amplitude of the pairing gap and roughly follows as $|\Delta(T)|^2$~\cite{Lutchyn2009,Taylor:13}, which is not a qualitative change and will thus not be discussed in detail. 

\begin{figure}[htbp]
    \centering
    \includegraphics[width=8.5cm]{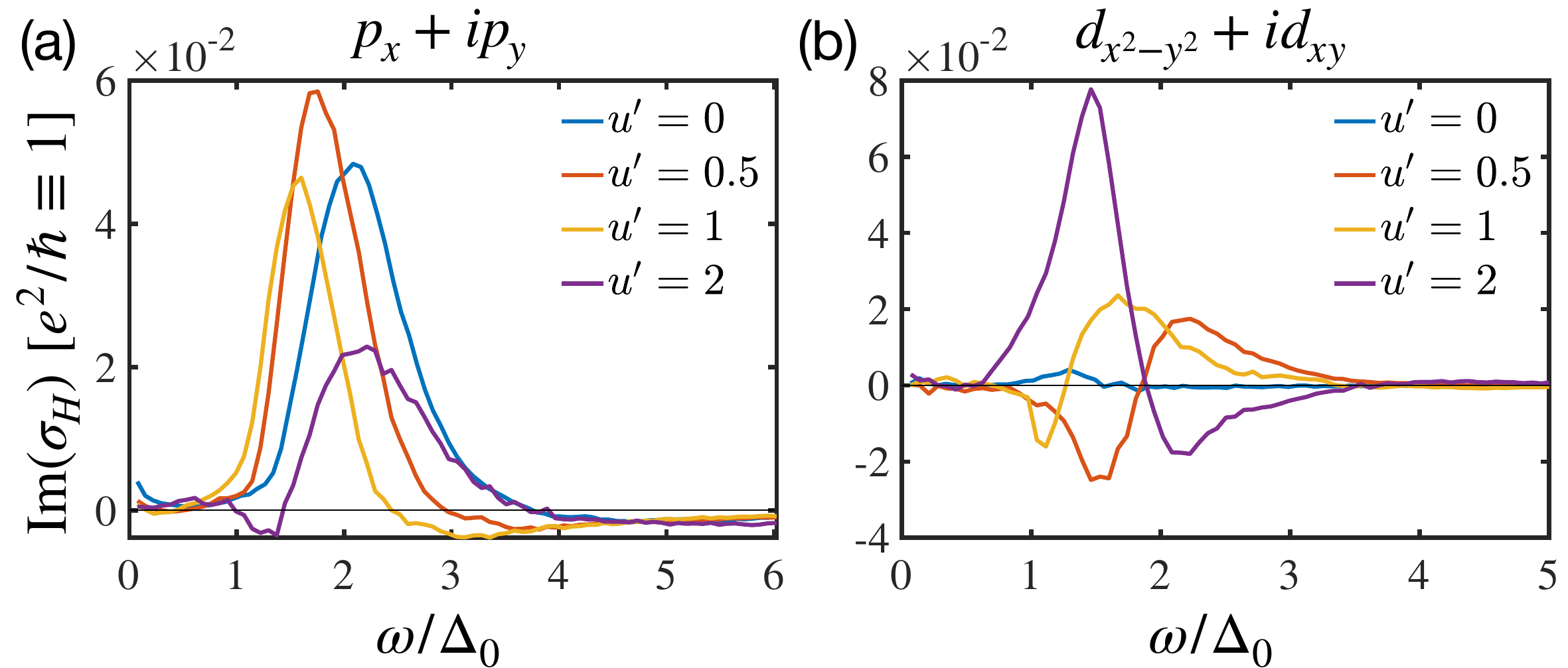}
    \caption{The imaginary part of the zero-temperature Hall conductivity of (a) $p_x+ip_y$ and (b) $d_{xy}+id_{x^2-y^2}$ states at $n_i=1.56\%$ and with different impurity potential profiles as determined by the potential $u^\prime$ assigned to the four nearest neighbors of each impurity site.}
    \label{fig2}
\end{figure}

\subsubsection{$p_x+ip_y$} 
The $p_x+ip_y$ state exhibits the strongest Hall effect. Our simulation roughly reproduces the general frequency-dependence of the Hall conductivity originally obtained in diagrammatic analyses~\cite{Lutchyn2009}, with peaks around $\omega = 2\Delta_0$ where $\Delta_0$ is the excitation gap extracted from the clean-limit density of states (DoS) distribution [Fig.~\ref{fig1} (c)]. However, some distinct features are new to us. As can be seen in Fig.~\ref{fig1} (a), the conductivity peaks are broadened. In particular,  $\text{Im}(\sigma_H)$ displays finite spectral intensity below $\omega=2\Delta_0$, unlike the diagrammatic result which is cut off from below at this frequency. Since $ \big[V^{mn}_{x}V_{y}^{nm}-(x\leftrightarrow y)\big]=2i\im (V^{mn}_{x,\bk}V_{y,\bk}^{nm})$ is purely imaginary, it is easy to check from (\ref{hallcond}) that $\text{Im}(\sigma_H)$ is associated with the quasiparticle excitation spectrum. Hence the finite intensity below $\omega =2\Delta_0$ can only be attributed to additional quasiparticle excitations below the superconducting gap, which were not captured in previous diagrammatic studies but naturally emerge in real-space simulations. Indeed, Fig.~\ref{fig1} (c) shows substantial gap-filling for samples with random impurities. Meanwhile, the softening of the coherence peak contributes in part to the broadening of the conductivity peaks. Additionally, the diagrammatic analyses revealed that the conductivity grows linearly with weak impurity concentration~~\cite{Goryo2008,Lutchyn2009}. Our calculation, as is shown in the inset of Fig.~\ref{fig1} (a), roughly reproduces this trend in the low concentration regime, although a deviation from this behavior at larger impurity concentration is also clearly observed. We finally note that similar behavior is anticipated for the two mixed helical non-unitary p-wave states proposed in Ref.~\onlinecite{Huang2021} --- which may be viewed as composites of $p_x+ip_y$ and $p_x-ip_y$ subsectors. 

\subsubsection{$d_{x^2-y^2}+id_{xy}$} 
Compared to $p_x+ip_y$, the $d_{x^2-y^2}+id_{xy}$ state in the presence of pointlike impurities appears to exhibit substantially smaller Hall conductivity, which reveals itself only after a statistical average over multiple impurity configurations [Fig.~\ref{fig1} (b)]. Notably, as we show in Appendix \ref{append1}, with increasing sample size, the error bars shrink, the curve of the sample average becomes smoother and its overall lineshape remains unchanged for any given impurity concentration. However, due to the limits in our computation resource, we cannot perform calculations with sufficiently large system size to bring the error bars to negligible levels. Hence future work is needed to unambiguously confirm the lineshape.  Meanwhile, unlike in the p-wave case, the majority of the weak conductivity spectrum lies below $\omega = 2\Delta_0$, suggesting a dominant contribution involving impurity-induced subgap states [Fig.~\ref{fig1} (d)]. This appears to agree with the finding in Li et.~al.~\cite{Li2020}, which demonstrated anomalous Hall effect associated with the impurity bands formed by impurity superlattices embedded in any 2D chiral superconductor. Nonetheless, Li et.~al.~\cite{Li2020}  did not contain sufficient
information about the relative strength of the Hall effect between p-wave and non-p-wave states.  

The strong suppression in chiral d-wave also roughly agrees with Goryo~\cite{Goryo2008}, where all non-p-wave chiral states displaying continuous symmetry were shown to have vanishing Hall conductivity at the level of pointlike skew-scattering diagrammatic analysis. The criterion is related to the azimuthal integral of $U_0^3\sin\theta_{\bk\bp}\sin l_z\theta_{\bk\bp}$~\cite{Goryo2008}, where a constant $U_0$ describes the $s$-wave scattering matrix associated with pointlike scatterers, and $l_z$ is the $z$-projection of the Cooper pair orbital angular momentum and $\theta_{\bk\bp}$ denotes the angle between wavevectors $\bk$ and $\bp$. For the $p_x+ip_y$ state, $l_z=1$, and the integral is finite. For the $d_{x^2-y^2}+id_{xy}$ state, $l_z=2$, and the integral vanishes. Note that, by simple extension, Goryo's conclusion shall also apply to the present chiral d-wave state on a square lattice. On the other hand, according to Ref.~\onlinecite{Ngampruetikorn:20}, non-pointlike impurities (or, finite-range impurities) can induce substantially enhanced Hall response in non-p-wave chiral states. The underlying physics can be understood as follows. While pointlike impurities induces only $s$-wave scatterings, non-pointlike impurities are able to generate higher angular momentum scatterings. Hence the constant $U_0$ is replaced by a generic impurity scattering matrix
\begin{equation}
U_{\bk\bp} = \sum_l U_{l}\cos l\theta_{\bk\bp} \,,
\end{equation} 
where $U_l$ denotes the strength of the scattering in the angular momentum $l$ channel. Following the skew-scattering analyses, the above expression for azimuthal integration is now replaced by 
\begin{equation}
U_{\bk\bk_1}U_{\bk_1\bp}U_{\bp\bk}\sin\theta_{\bk\bp}\sin l_z\theta_{\bk\bp} \,,
\end{equation} 
where $\bk_1$ is an intermediate wavevector which must also be summed over. One can always find skew-scattering processes which ensure that the above expression integrates to a finite value, thereby ensuring finite Hall conductivity. One simple example is when the first two scattering events occur in the $l=0$ ($s$-wave) channel and the third one in the $l=l_z-1$ channel. Then the expression becomes $U_0^2U_{l_z-1}\cos(l_z-1)\theta_{\bk\bp}\sin\theta_{\bk\bp}\sin l_z\theta_{\bk\bp}$, whose azimuthal integral is nonzero. 

In fact, the same argument also explains the finite albeit minuscule Hall conductivity in our chiral d-wave model with pointlike impurities. The key is to view clusters of nearby pointlike impurities as effective finite-radius impurities. This is thus a distinct effect only observable in real-space calculations, absent in momentum-space analyses~\cite{Goryo2008,Lutchyn2009,Ngampruetikorn:20}. Understandably, this effect is weak in the case of dilute impurities. Below, we turn to the scenario where the individual impurities are non-pointlike. 

As a simple simulation of non-pointlike impurities with finite range potential profile, we add extra potential $u'$ to each point-impurity's four neighboring sites. The results for both p-wave and d-wave states are shown in Fig.~\ref{fig2} for comparison. As one can see, while the magnitude of the p-wave Hall conductivity does not change qualitatively with varying $u'$, the d-wave conductivity readily gains magnitude similar to that of the p-wave state once a finite $u'$ is turned on. This confirms our expectation and is consistent with Ref.~\onlinecite{Ngampruetikorn:20}. Note that the considerable change of the conductivity lineshape with different $u^\prime$ is related to the variation of the subgap state energy spectrum.

\subsubsection{Non-chiral states} 
Finally, in agreement with Goryo~\cite{Goryo2008}, our calculations did not reveal any finite Hall conductivity in neither $s+i d_{x^2-y^2}$ nor $d_{x^2-y^2}+i g_{xy(x^2-y^2)}$ state. Nonetheless, both conductivity exhibits clear statistical fluctuations around zero (the Hall conductivity of the former is shown in Fig.~\ref{app1} in Appendix~\ref{append1}, while that of the latter is not shown). This can be ascribed to the existence of scattering events that contribute with opposite signs to the Hall conductivity. We expect the same conclusion to hold for other mixed-representation non-chiral states, such as $s+id_{xy}$.

\subsection{Self-consistent calculations}
The above non-self-consistent calculations did not account for the impurity-induced local order parameter variations. This in part resembles the scenario in the momentum-space analyses~\cite{Goryo2008,Lutchyn2009,Konig:17,Ngampruetikorn:20} where the disorder effects on the order parameters can at best be treated in an average manner. However, even with pointlike impurities, the spatial variation of the order parameters typically spans over a coherence length and may therefore enhance non-s-wave scatterings. This could have significant implications for the Hall effect in non-p-wave chiral states, as we verify below through self-consistent calculations. 

The superconducting order parameters on the lattice bonds (or sites) are self-consistently determined by 
\begin{align}\label{gap_selfc}
    \Delta_{\bi\bj}&=\widetilde{U}_{\bi\bj}\braket{\hc_{\bi}\hc_{\bj}} \,,
\end{align}
where the spin indices is omitted, $\widetilde{U}_{\bi\bj}$ is the effective interaction to create Cooper pairs between electrons at the $\bi$-th and the $\bj$-th sites, and $\braket{\cdots}$ denotes the expectation value of the ground state. To obtain a target order parameter amplitude $\Delta_{0,\bi\bj}=\Delta$ in the clean limit, we first determine the corresponding interaction strength via $\widetilde{U}_{\bi\bj} = \Delta/\braket{\hc_{\bi}\hc_{\bj}}$, where $\braket{\hc_{\bi}\hc_{\bj}}$ is evaluated based on the BdG Hamiltonian with initial pairing $\Delta$. In the presence of random impurities, using an initial gap amplitude $\Delta_{0,\bi\bj}$ and the interaction $\widetilde{U}_{\bi\bj}$, we diagonalize the full BdG Hamiltonian $\hH_{\bdg}=\hH_{\mathrm{TB}}+\hat{\Delta}+\hH_{\imp}$ and the Eq.~\eqref{gap_selfc} will give new gap amplitude at each bond. Using the new gap amplitude, we diagonalize the full BdG Hamiltonian and plug all parameters into Eq.~\eqref{gap_selfc} again. With the initial gap amplitude $\Delta_0\sim 0.1t$, we repeat this process until the change of gap amplitude between two successive iterations is smaller than $10^{-3}t$. The results below are based on self-consistent calculations of models with lattice size $50\times 50$. 

\begin{figure}[htbp]
    \centering
    \includegraphics[width=8.5cm]{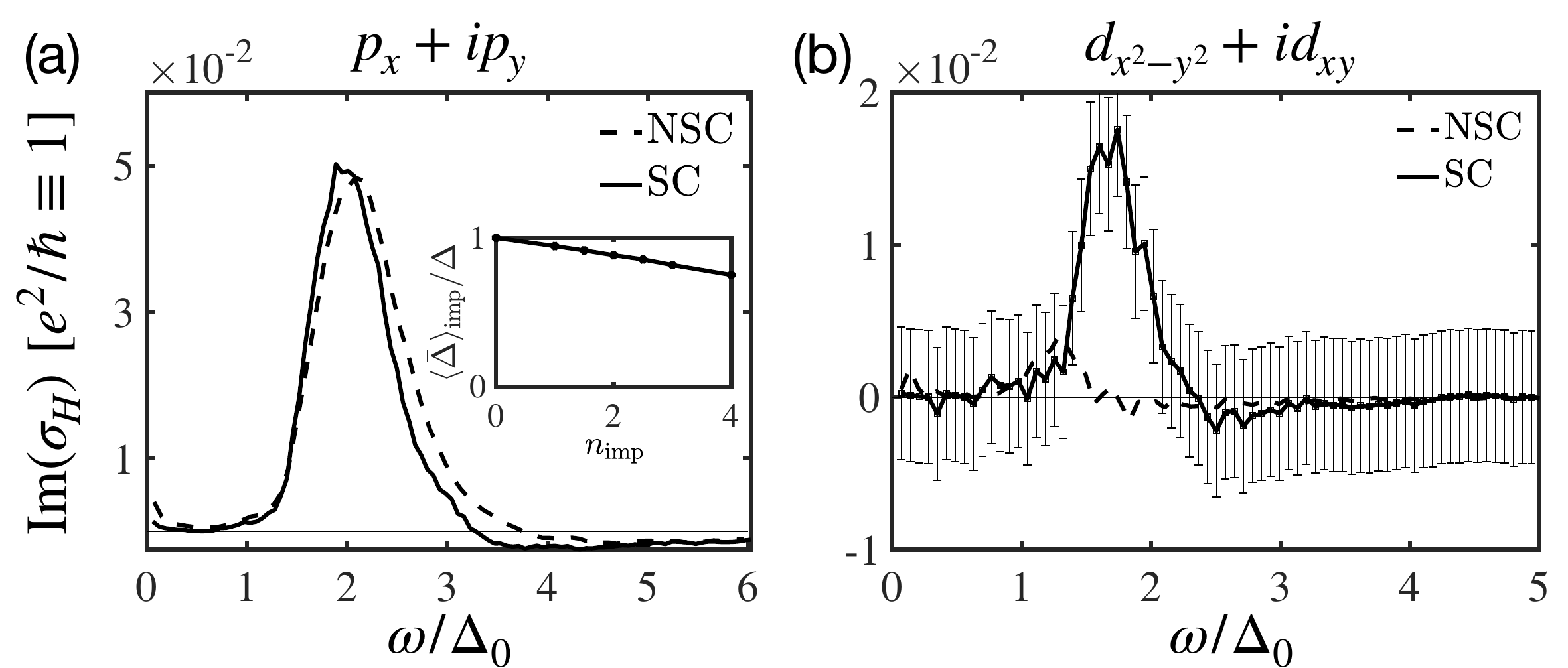}
    \caption{The imaginary part of the zero-temperature Hall conductivity of (a) $p_x+ip_y$ and (b) $d_{xy}+id_{x^2-y^2}$ states for pointlike impurities at the impurity concentration $n_{\mathrm{imp}}\simeq 1.5\%$ after self-consistency. Parameters are the same as in Fig.~\ref{fig1}. The dash curves are the same as the Im[$\sigma_H$] in Fig.~\ref{fig1} (a) and (b). The abbreviations in the legend NSC and SC denote non-self-consistent calculation and self-consistent calculation, respectively. The inset in (a) shows the drop of the spatial average of the pairing order parameter as a function of the impurity concentration. The error bars in the self-consistent data in (b) are to demonstrate the robustness of the lineshape. }
    \label{fig3}
\end{figure}

To characterize the disorder-induced suppression of superconductivity, we evaluate the spatial average of the pairing order parameters on certain lattice bonds, $\braket{\Bar{\Delta}}_{\imp}=\frac{1}{N}\sum_{\bi}\braket{\Delta_{\bi+\bfdelta,\bi}}_{\imp}$. The result for the $p_x+ip_y$ state is shown in the inset of Fig. \ref{fig3} (a), where it can be seen that superconductivity is robust up to the maximal impurity concentration $n_{\imp}= 4\%$ employed here. The $d_{x^2-y^2}+id_{xy}$ state exhibits similar dependence on the impurity concentration.

Figure \ref{fig3} presents the imaginary part of the Hall conductivity for both $p_x+ip_y$ and $d_{x^2-y^2}+id_{xy}$ states, evaluated using self-consistent order parameter profiles in lattices with random pointlike impurities. The results of the corresponding non-self-consistent calculations are also shown for comparison. As one can see from Fig.~\ref{fig3} (a), the p-wave state is essentially unchanged upon order parameter self-consistency. However, self-consistency drastically elevates the $d$-wave Hall conductivity (Fig.~\ref{fig3} (b)) to a magnitude comparable to that in p-wave as well as to those induced by finite-radius impurities (Fig.~\ref{fig2} (b)). This indicates much enhanced non-$s$-wave scattering as a result of the spatially inhomogeneous order parameters. Our result therefore highlights the importance to take into account the order parameter spatial variation -- which cannot be captured in momentum-space analyses -- when studying the Hall effect of non-p-wave chiral states. Finally, we also checked that non-chiral states cannot generate finite Hall conductivity even with self-consistency.

\section{Two-band models}
\label{sec:twoband}
Multiband models possess additional intrinsic contribution to the anomalous Hall conductivity, which is related to virtual interband optical transitions, such as $|m,\bk\rangle \rightarrow |n,\bk\rangle \rightarrow |m,\bk\rangle $ with $\varepsilon_{m,\bk}<0<\varepsilon_{n,\bk}$ and $|\varepsilon_{m,\bk}|\ne |\varepsilon_{n,\bk}|$. The aim of this section is to investigate the fate of this intrinsic contribution against random disorder. For illustration, we consider the $p_x+ip_y$ state on a two-orbital (two-band) model with $d_{xz}$ and $d_{yz}$ orbitals residing on each site of a square lattice. The normal state of the model is described by the following tight-binding Hamiltonian:
\begin{align}
    &\hH_{\mathrm{TB}}=-\mu\sum_{\bi}\had_{\bi}\ha_{\bi}-t\sum_{\bi}\left(\had_{(i_x,i_y+1)}\ha_{\bi}+\HC\right)\non\\
    &-\mu\sum_{\bi}\hbd_{\bi}\hb_{\bi}-t\sum_{\bi}\left(\hbd_{(i_x+1,i_y)}\hb_{\bi}+\HC\right)\non\\
    &+t'\sum_{\bi}\left(\hbd_{(i_x+1,i_y+1)}\ha_{\bi}-\hbd_{(i_x-1,i_y+1)}\ha_{\bi}+\HC\right)
\end{align}
where $\had/\ha$ and $\hbd/\hb$ are creation/annihilation operators for $d_{xz}$ and $d_{yz}$ orbitals, respectively, and $t^\prime$ denotes the interorbital hybridization. Note that spin indices have been dropped for simplicity. Without loss of generality, it is assumed that the two orbitals feel equivalent impurity potential, so that impurity Hamiltonian is given by $\hH_{\imp}=\sum_{\bi}u(\had_{\bi}\ha_{\bi}+\hbd_{\bi}\hb_{\bi})$. The $p_x+ip_y$ pairing acquires the following momentum-space form $\hH_{\Delta}=\sum_{\bk}\Delta_0\big(\sin k_x\had_{\bk\ua}\had_{-\bk\da}+i\sin k_y \hbd_{\bk\ua}\hbd_{-\bk\da}+\HC\big)$.

\begin{figure}
    \centering
    \includegraphics[width=8.5cm]{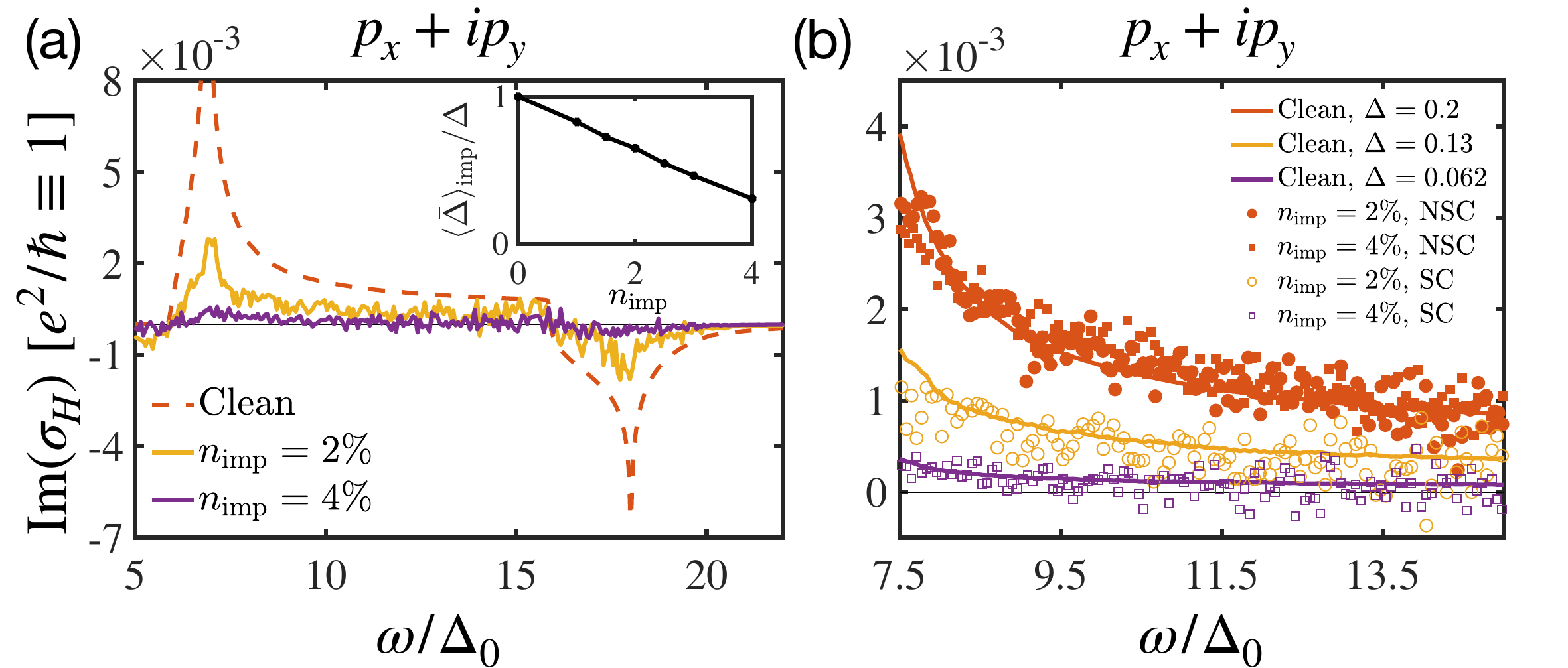}
    \caption{(a) The imaginary part of Hall conductivity in the two-band model. The results for finite impurity concentration scenarios are obtained after order parameter self-consistency. The inset shows the drop of the average order parameter potential with varying impurity concentration. (b) Zoom-in view of part of the high frequency Hall conductivity. The solid curves plot the clean-limit Hall conductivity evaluated using the pairing amplitude $\Delta=\Delta_0$ as indicated for each curve. The filled circles and squares depicted the conductivity of disordered samples in non-self-consistent (NSC) calculations with uniform input pairing amplitude $\Delta_0=0.2t$. The open circles and squares plot the conductivity obtained in self-consistent (SC) calculations, which use an interaction that produces a pairing amplitude of $\Delta_0=0.2t$ in the clean limit. These calculations employed the parameter set $(t^\prime,\mu)=(0.5,1)t$.}
    \label{fig4}
\end{figure}

\begin{figure}
    \includegraphics[width=6cm]{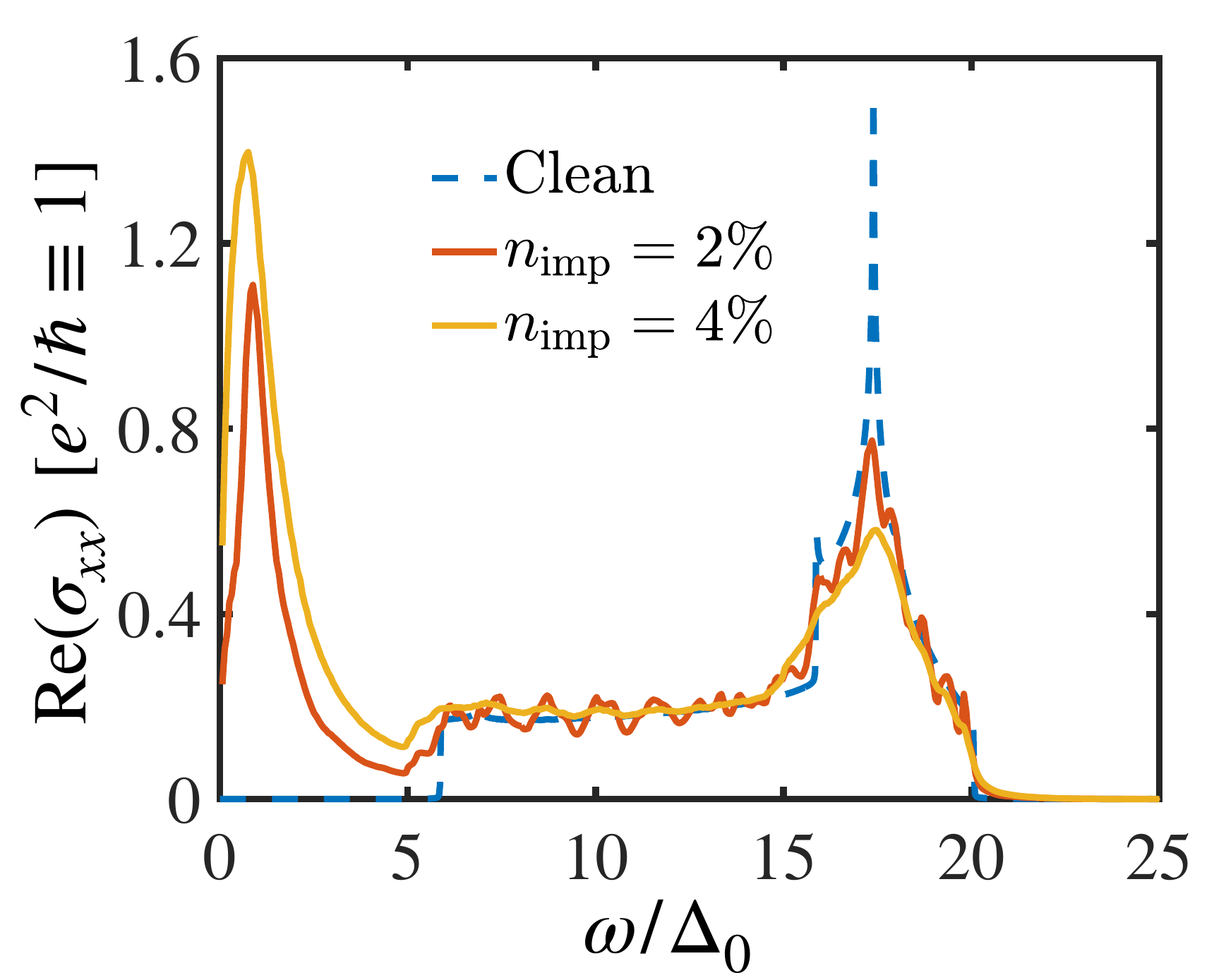}
    \caption{The real part of longitudinal conductivity $\sigma_{xx}$ for the two-band $p_x+ip_y$ model in clean and disordered systems. The results for disordered systems are obtained from the same models as in Fig.~\ref{fig4}. The clean-limit conductivity (blue dashed curve) was evaluated using the momentum-space formulation  provided in the text. The Drude peak in $\sigma_{xx}$ is suppressed by superconductivity. Note that the oscillatory high-frequency conductivity is a numerical artifact of our finite-size modeling. It is more severe in smaller system size calculations and at lower impurity concentrations, where the energy distribution of the limited number of high-energy quasiparticle states is not yet sufficiently spread out in statistical sense. }
    \label{fig5}
\end{figure}

The above model has often been used in the effective description of the multiband superconductivity in \SRO~driven by its Ru $d_{xz}$ and $d_{yz}$ orbitals~\cite{Kivelson2010,Kallin2012}. In the present study, we intentionally choose a large $t^\prime$ to ensure a sizable band separation at generic momenta, which then allows us to differentiate between the conductivity arising respectively from interband and intraband processes. To be more specific, while the intraband transitions dominate the imaginary part of the Hall conductivity around $\omega = 2\Delta_0$ as we already saw in the previous section (except for some subtle multiband effects here), the interband conductivity emerges above $\omega \sim \Delta\varepsilon+\Delta_0$, where $\Delta\varepsilon$ is the band separation energy at the Fermi wavevector and is determined by $t'$. The numerical calculations were done for lattices of size $60\times 60$ and we take at least $30$ samples for impurity ensemble average. Figure \ref{fig4} (a) presents some representative results obtained from our self-consistent calculation under the influence of random disorder (real part of $\sigma_H$ not shown). The inset of Fig. \ref{fig4} (a) shows a the drop of the order parameter as a function of the impurity concentration, which exhibits suppressed but still robust superconductivity up to $n_\text{imp} =4\%$.

%While not the focus of this section, the low-frequency Hall conductivity in this particular model differs from the typical single-band behavior shown in the preceding section. For one, the conductivity peak is located below $\omega \approx 2\Delta$ and it shifts with impurity concentration. Furthermore, the peak value decreases with increasing disorder. 

Let's focused on the high frequency conductivity arising from interband contributions. The dashed curve in Fig.~\ref{fig4} (a) plots the clean-limit intrinsic Hall conductivity. For comparison, the dashed curve in Fig.~\ref{fig5} (b) shows the clean-limit intrinsic longitudinal conductivity $\sigma_{xx}(w)$ generated by similar interband transitions, evaluated according to the formula,
 \begin{equation}\label{drudecond}
     \sigma_{xx}(\omega)=\frac{i}{2N\omega}\sum_{\bk,m,n}\frac{|V_{x,\bk}^{mn}|^2[f(E_{m,\bk})-f(E_{n,\bk})]}{\omega+i\eta-E_{n,\bk}+E_{m,\bk}} \,.
 \end{equation}
And following the same method in the section \ref{sec:formalism}, the longitudinal condcutivity for disordered system is now evaluated according to the real-space current-current correlation similar to Eq.~\eqref{hallcond}, except that the form of $V^{mn}_{x}V_{y}^{nm}-(x\leftrightarrow y)$ there is now replaced by $|V_x^{mn}|^2$ and the factor 1/4 is now corrected by 1/2 due to the definition of longitudinal conductivity. Both conductivities are cut off at a similar frequency well above $\omega=2\Delta_0$. 

Upon the introduction of impurities, the intrinsic Hall conductivity across a wide interband frequency window is noticeably suppressed, with increasing suppression as the impurity concentration increases [Fig.~\ref{fig4} (a)]. Since the Hall effect originates from the chiral Cooper pairing, such level of suppression is ascribable to the parametric disorder-suppression of superconductivity. On the one hand, away from the peaks between the frequency interval $7.5\Delta_0\sim 15\Delta_0$, the non-self-consistent Hall conductivity (the filled circles and squares in Fig.~\ref{fig4} (b)) is barely influenced by varying degree of disorder. On the other hand, the self-consistent Hall conductivity can be seen to roughly track the clean-limit results obtained by using uniform order parameters the same magnitude as the corresponding $\braket{\bar{\Delta}_{\bi\bj}}_\text{imp}$ (Fig.~\ref{fig4} (b)). Such strong parametric dependence of the self-consistent Hall response on $\braket{\bar{\Delta}_{\bi\bj}}_\text{imp}$ also agrees with expectation that the intrinsic Hall conductivity shall be proportional to $\Delta^2$~\cite{Kallin2012,Taylor:13}. By contrast, the intrinsic longitudinal conductivity is almost unaffected by the same degree of disorder (Fig.~\ref{fig5}), as it does not rely on having superconducting pairing.

Note that intrinsic Hall effect has also been demonstrated for multiband superconductors with higher angular momentum chiral pairing, such as chiral $d$-wave and chiral $f$-wave~\cite{Kallin2017,Brydon2019,Zhang2020}. We expect similar strong disorder suppression of interband Hall conductivity in those scenarios.

\section{summary and final remarks}
\label{sec:remarks}
We have studied the effects of impurity scatterings on the anomalous Hall effect in a number of TRSB superconducting states, on the basis of real-space simulations of lattice models with random impurities. Our calculations of the single-band chiral p-wave model reproduce the qualitative behavior previously obtained in diagrammatic studies, but also show some quantitative difference due to the presence of impurity-induced subgap quasiparticle excitations not considered previously. Further, in contrast to the diagrammatic analysis, we demonstrated anomalous Hall response in non-p-wave chiral states, which is minuscule in non-self-consistent calculations of models with pointlike impurities but readily becomes substantial with finite-range impurities or with self-consistently resolved inhomogeneous superconducting order parameter. In addition, we verified that random impurities do not induce Hall effect in non-chiral TRSB superconductors. Our study highlights the importance of taking into account the order parameter inhomogeneity when studying the Hall response of chiral superconductors.

In the two-band chiral superconducting model, the high-frequency intrinsic Hall conductivity originating from interband optical transitions was found to depend strongly on the impurity scatterings, roughly following the parametric disorder-suppression of the superconductivity. By contrast, the longitudinal conductivity of similar interband origin remains robust against the same degree of disorder. 

Finally, our results may have some meaningful implication for the polar Kerr measurement and the differentiation of various chiral superconducting states. Particularly relevant is the recognizably different impurity-concentration dependence of the low-frequency (i.e. around $\omega \sim 2\Delta$) and high-frequency (i.e. $\omega \sim$ band separation energy scales) Hall conductivity. By controlling the impurity concentration in the sample material, the impurity effects on the Kerr rotation angle could thus be analyzed with some level of confidence if the probing photon energy lies within certain frequency range. Meanwhile, our study also shows that Kerr effect is absent in non-chiral but TRSB superconducting states even in the presence of random impurities. This would cast some doubt on the recent proposals of $s+id$ and $d+ig$ pairings in \SRO, such as in Refs.~\onlinecite{Anderson2019,Kivelson2020,Ghosh2021}. Finally, while our study has been conducted with \SRO~at hand, our conclusions shall also hold for general TRSB superconductors. 

\section{Acknowledgements} 
We acknowledge helpful discussions with Catherine Kallin and Jia-Long Zhang. This work is supported by NSFC under grant No.~11904155, the Guangdong Provincial Key Laboratory under Grant No.~2019B121203002, the Guangdong Science and Technology Department under Grant 2022A1515011948, and a Shenzhen Science and Technology Program (Grant No. KQTD20200820113010023). Computing resources are provided by the Center for Computational Science and Engineering at Southern University of Science and Technology.

\appendix
\section{}
\label{append1}
\begin{figure}[h]
    \centering
    \includegraphics[width=8.5cm]{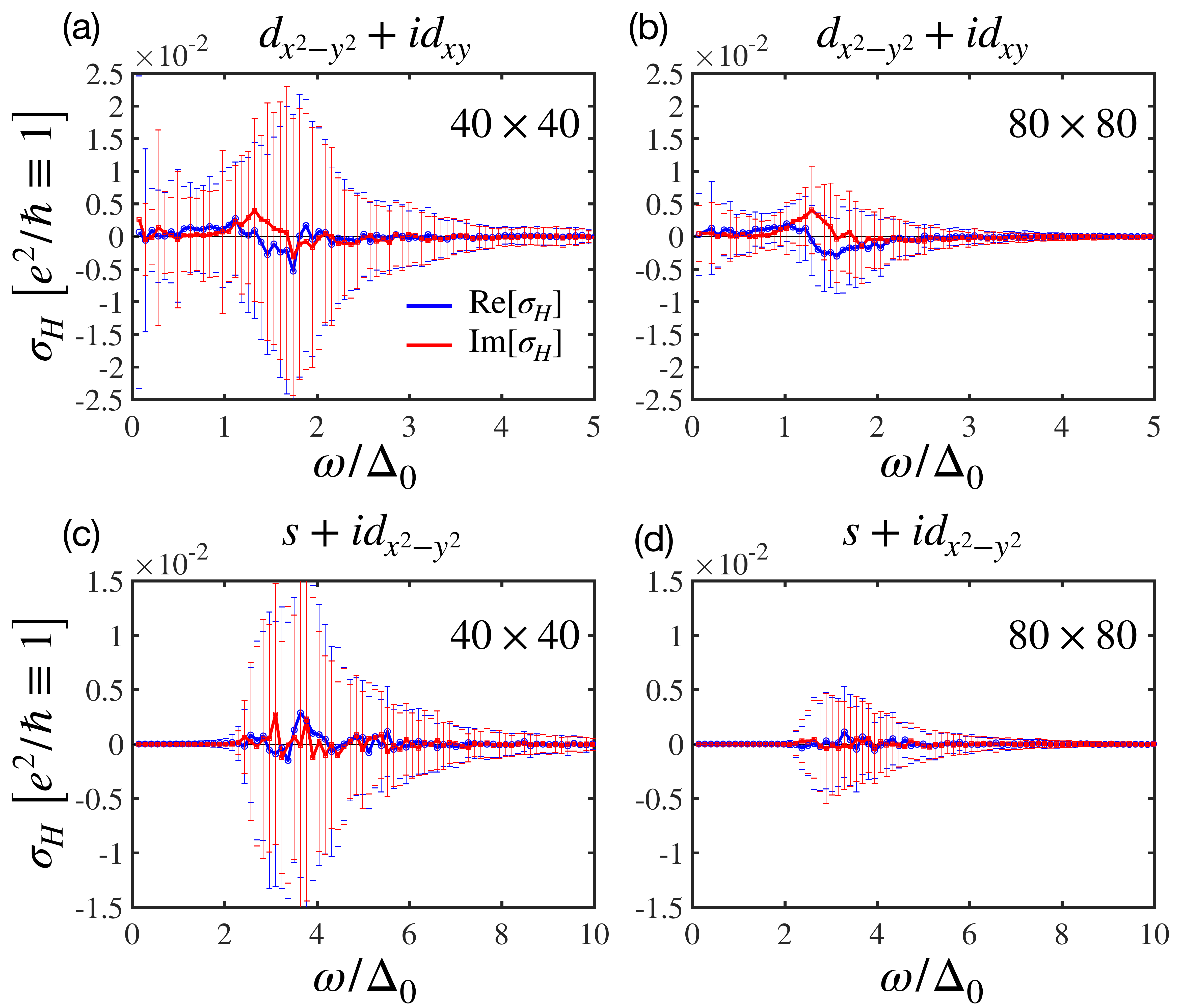}
    \caption{Comparison of the Hall conductivity obtained from calculations with respective system size of $40 \times 40$ and $80 \times 80$ in the presence of the same concentration of random pointlike impurities, for the single-band chiral d-wave (upper panel) and $s+id_{x^2-y^2}$ (lower panel) states. Except for the different system size, all other parameters are the same as in Fig. \ref{fig1} (b) for the chiral d-wave model, and $\Delta_0=0.09t$ for the $s+id_{x^2-y^2}$ model. All data sets are obtained by averaging over 90 different impurity samples.}
    \label{app1}
\end{figure}

In this appendix, we compare the results of some single-band calculations with two different system sizes but with the same concentration of random pointlike impurities. For the single-band chiral d-wave state, as one can see from Fig.~\ref{app1} (a) and (b), the scatter of the data (measured by the size of the error bars) decreases noticeably with increasing system size, and the overall lineshape of the two curves appears to become smoother. However, we are unable to perform calculations with sufficiently large system size that could reduce the error bars to negligible levels. Hence further calculation is needed to unambiguously confirm the d-wave Hall conductivity lineshape in the case of pointlike impurities. Also shown in Fig.~\ref{app1} are the results of the single-band $s+id_{x^2-y^2}$ model. In contrast to the chiral d-wave state, the $s+id_{x^2-y^2}$ conductivity revolves around zero and does not exhibit any stable lineshape as the system size is varied. This indicates vanishing Hall response in this state. 

\begin{thebibliography}{0}
\bibitem{footnote} Here, we are referring to the Hall conductivity by taking the limit of zero wavevector $\bq \rightarrow 0$ and finite frequency $\omega \neq 0$. Non-zero conductivity can in fact be obtained for finite $\bq$. However, the single-band finite-$\bq$ conductivity estimated based on the actual experimental conditions in Kerr measurement is orders of magnitude smaller than needed to explain the observation.
\end{thebibliography}
\end{document}